\documentclass[conference]{IEEEtran}

\usepackage{graphicx}
\usepackage{caption}
\usepackage{amsmath}
\usepackage{multirow}

\usepackage{subfig}

\usepackage{url}
\usepackage{float}
\usepackage{color}
\usepackage{enumerate}

\newcommand{\ignore}[1]{}

\newcommand{\ya}{\textsc{\emph{YA}}}

  \makeatletter
  \let\@copyrightspace\relax
  \makeatother

\begin{document}

%\title{The Social Ecosystem of Yahoo! Answers: Friends, Activities and Abuses}
%\title{Deviance and Retention in Yahoo! Answers}

\title{Privacy Concerns vs. User Behavior \\in Community Question Answering \\ { \large [Please cite the ASONAM'15 version of this paper]}}
%\subtitle{[Please cite the ASONAM'15 version of this paper]}

\author{
\IEEEauthorblockN{Imrul Kayes}
\IEEEauthorblockA{
%Computer Science and Engineering\\
University of South Florida\\
Tampa FL, USA\\
imrul@mail.usf.edu}
\and
\IEEEauthorblockN{Nicolas Kourtellis}
\IEEEauthorblockA{
Telefonica Research\\
Barcelona, Spain\\
nicolas.kourtellis@telefonica.com}
\and
\IEEEauthorblockN{Francesco Bonchi}
\IEEEauthorblockA{
Yahoo Labs\\
Barcelona, Spain\\
bonchi@yahoo-inc.com}
\and
\IEEEauthorblockN{Adriana Iamnitchi}
\IEEEauthorblockA{
%Computer Science and Engineering\\
University of South Florida\\
Tampa FL, USA\\
anda@cse.usf.edu}
}

\maketitle

\begin{abstract}

Community-based question answering (CQA) platforms are crowd-sourced services for sharing user expertise on various topics, from mechanical repairs to parenting. 
While they naturally build-in an online social network infrastructure, they carry a very different purpose from Facebook-like social networks, where users ``hang-out'' with their friends and tend to share more personal information. 
It is unclear, thus, how the privacy concerns and their correlation with user behavior in an online social network translate into a CQA platform. 
This study analyzes one year of recorded traces from a mature CQA platform to understand the association between users' privacy concerns as manifested by their account settings and their activity in the platform.
%This study attempts to understand the contribution behavior of the privacy concerned users in a mature CQA.
The results show that privacy preference is correlated with behavior in the community in terms of engagement, retention, accomplishments and deviance from the norm. 
We find privacy-concerned users have higher qualitative and quantitative contributions, show higher retention, report more abuses, have higher perception on answer quality and have larger social circles.
However, at the same time, these users also exhibit more deviant behavior than the users with public profiles.
%Using the quantitative behavioral differences,  we build a classifier that is able to  predict users'  privacy with an accuracy as high as 75\%.

%Our results show that a user's privacy preference is  related to his  contribution behavior characterized by a number of metrics such as engagement, retention, accomplishment and deviance. 
%We also find that users' privacy preferences are culturally determined:  on average users from individualistic cultures (e.g., US) are more privacy concerned than collectivist cultures (e.g., Mexico). 
%Based on our empirical observations, we build a classifier that is able to  predict users'  privacy with an accuracy as high as 75\%.

%In addition to better understanding the user behavior in such platforms, this study suggests ways to 

%\category{K.4.2}{Computers and Society}{Social Issues---Abuse and crime
%involving computers}
%A category including the fourth, optional field follows...
 %[Computers and Society]: Social IssuesÑAbuse and crime
%involving computers;
%\category{J.4}{Social and Behavioral Sciences}{Sociology}

%\terms{Measurement, Human Factors}

%\keywords{Community question answering; privacy concerns; crowdsourcing;}

\end{abstract}


\section{Introduction}\label{sec:intro}

Community-based Question-Answering (CQA) platforms, such as Yahoo Answers (\ya), Quora and Stack Overflow, are online platforms where community members ask and answer questions. 
%These sites are now working as rich and mature repositories of user-contributed questions and answers. 
For example, \ya, launched in December 2005, has more than one billion posted answers~\cite{YahooAnswersBlog}.
Liu et al.~\cite{Liu2012WSF} found that about $2$\% of web searches performed by users of \ya\ lead to a question posted to the community.

Such communities have a social network component, where users can follow other users' activity via updates.
Privacy settings are typically available for users to personalize. 
Two conflicting goals in privacy setting configurations emerge: on one hand, the platform is most useful when user-generated content is publicly available. 
On the other hand, various studies on general-purpose online social networks (such as Facebook) showed that the users who exercise their privacy rights (specifically, by restricting the visibility of their content) are more engaged and thus contribute more to the community. 

To the best of our knowledge, this is the first study of the association between users' privacy concerns and contribution behavior in CQA platforms via analysis of user activity logs.
In our previous work on cultures in \ya~\cite{kayes2015culture}, we found that users' privacy concerns vary across cultures: users from individualistic countries are more concerned about their privacy compared to collectivistic countries.
However, this work doesn't explore the relationship  between users' privacy concerns and their contribution behavior.
In this work, we analyzed more than a year of activity traces from 1.5 million users from \ya\ to answer the following questions related to users' contribution behavior:\\
(1)~Are there quantitative and qualitative differences in user contributions between user groups with private vs. public settings?\\
(2)~Is user engagement (measured by frequency of contributing content and number of social contacts) correlated with user privacy settings?\\
(3)~Do users with privacy settings enabled tend to violate community norms more than users with public content?

Our study makes two main contributions.
First, while the previous related studies~\cite{boyd2010FB, Rainie2012Pew, Staddon2012PCT} on Facebook were based on self-reported data (shown to be subject to bias~\cite{gosling1998people,nishioka2012producing}), this study uses modifications of the privacy settings as a proxy of privacy concern, and users' recorded activity logs to infer their behavior.
This is the first data-driven study that %confirms the previous user survey-based 
shows correlation between privacy controls and online user behavior. 
Second, this study is the first that characterizes a CQA platform from the privacy perspective.
Our study finds that privacy-concerned users contribute more to the community. 
They are more engaged, having higher retention and larger social circles, and have higher perception on answer quality.
However, they also exhibit more violations of platform rules in asking and answering questions than the users with public profiles.

The paper overviews related work in Section~\ref{sec:related}, describes the \ya\ platform and our dataset in Section~\ref{sec:ya-details-datasets} and presents our data analysis in Section~\ref{sec:privacy}. 
%Section~\ref{sec:classification} shows the classification models for privacy settings.
We conclude with a discussion of results in Section~\ref{sec:discussion}.

\section{Related Work}\label{sec:related}

Community-based Question Answering has attracted much research interest from diverse communities as web science, HCI and information retrieval. 
We divide research on CQA in four categories: content perspective, user perspective, system perspective and social network perspective.
Content perspective research focuses on various aspects of questions and answers such as answerability of questions~\cite{dror2013will,Richardson2011Answerability}, question classification (e.g., factual or conversational)~\cite{Harper2009Type,harper2010question}, quality of questions~\cite{Li2012Question,Sun2009Questions} and answers~\cite{Agichtein2008Answers,Shah2010Answers}.
Kucuktunc et al.~\cite{Kucuktunc2012Sentiment} investigate the influence of gender, age, education level, and topic  on sentiments of questions and answers. 
%One aspect is the quality of questions~\cite{Li2012Question,Sun2009Questions} and answers~\cite{Shah2010Answers,Agichtein2008Answers,Jeon2006Answers}.
%Some studies~\cite{dror2013will,Richardson2011Answerability} attempted to assess the answerability of questions.
%Kucuktunc et al.~\cite{Kucuktunc2012Sentiment} investigated the influence of gender, age, education level, and topic  on sentiments of questions and answers. 
%Other studies~\cite{Harper2009Type,harper2010question} focused on classifying question (e.g., factual or conversational).

User perspective research sheds light on why users contribute content: that is, why users ask questions (askers are failed searchers, in that, they use CQA sites when web search fails~\cite{Liu2012WSF}) and  why they answer questions (e.g., they refrain from answering sensitive questions to avoid being reported for abuse and potentially lose access to the community~\cite{Dearman2010Why}).
Moreover, Liu et al.~\cite{liu2011modeling} explore the factors that influence users' answering behavior in \ya\ (e.g., when users tend to answer and how they choose questions).
Pelleg et al.~\cite{pelleg2012can} investigate truthfulness of users and offer a quantitative proof that users post sensitive and accurate information to fulfill specific information needs.
%Dearman et al.~\cite{Dearman2010Why} asked why users on YA do not answer questions and found that active answerers (who contribute most of the answers) do not want to get reported for abuse and potentially lose access to the community.
%Liu et al.~\cite{liu2011modeling} explored the factors that influence users' answering behavior in YA (e.g., when users tend to answer and how they choose questions).
%Liu et al.~\cite{Liu2012WSF} also found that a vast majority of the askers are failed searchers: when web search fails they become YA askers.
%Pelleg et al.~\cite{pelleg2012can} investigated truthfulness of users and offered a quantitative proof that users post sensitive and accurate information to fulfill specific information needs.

System perspective research develops techniques and tools to improve platform usability.
It includes routing questions to expert users~\cite{Qu2009PQR,Szpektor2013Reco}, extracting factual answers from QA archives~\cite{Bian2008Finding} and reusing the repository of past answers to answer new open questions~\cite{Shtok2012LPA}.
Weber et al.~\cite{Weber2012ALE} derive ``tips'' (a self-contained bit of non-obvious answer) from \ya\ to address ``how-to'' queries.
%One line of research focuses on question recommendation systems~\cite{Qu2009PQR,Szpektor2013Reco}.
%In~\cite{Bian2008Finding}, the authors described a method to retrieve factual answers from QA archives.
%Shtok et al.~\cite{Shtok2012LPA} attempted to reuse the repository of past answers to answer new open questions in order to reduce unanswered questions.
%Weber et al.~\cite{Weber2012ALE} derived ``tips'' (a self-contained bit of non-obvious answer) from YA to address ``how-to'' queries.
Social network perspective research attempts to understand the interplay between users' social connections and Q\&A activities such as analyzing  the social network of Quora~\cite{Wang2013WSC}, using social network properties and contribution behavior for  content abusers detection~\cite{Kayes2015WWW}.
%Wang et al.~\cite{Wang2013WSC} analyzed the social network of Quora and found that users who contributed more and better answers tend to have more followers.
%Kayes et al.~\cite{Kayes2015WWW} detected content abusers on YA based on their social network properties and contribution behavior.

A number of studies~\cite{boyd2010FB, Rainie2012Pew, Staddon2012PCT} on social networks like Facebook have shown the correlation between users' self-reported privacy concerns and their self-reported behavior.
For example, Staddon et al.~\cite{Staddon2012PCT} showed that 
users who express concerns on Facebook privacy controls and find it difficult to comprehend sharing practices also report less engagement such as visiting, commenting, and liking content.
At the same time, users who report more control and comprehension of privacy settings and their consequences are more engaged with the platform.
%Another user survey-based study~\cite{boyd2010FB} on Facebook finds that %while young adults tend to not change the default privacy settings, 
Similarly, the frequency of visits, type of use, and general Internet skills are shown to be related to the personalization of the default privacy settings~\cite{boyd2010FB}. 
Acquisti and Gross'~\cite{Acquisti2006Imagined} survey on Facebook finds that a user's privacy concerns are only a weak predictor of his joining the network: that is, despite expressing privacy concerns, users join the network and reveal great amounts of personal information.
Young et al.~\cite{Young2009FB} used surveys and interviews on Facebook users to show that Internet privacy concerns and information revelation are negatively correlated.
Tufekci's study~\cite{Tufekci2008Can} on a small sample (704) of college students shows that students on Facebook and Myspace manage privacy concerns by adjusting profile visibility but not by restricting the profile information.

Wang et al.'s ~\cite{wang2011concerned} demographic study on privacy concerns among American, Chinese, and Indian social network users shows that American respondents are the most privacy concerned, followed by Chinese. % and Indians. 
However, there has been no research on privacy concerns and user behavior in CQA platforms.
Our previous work~\cite{kayes2015culture} on cultures in \ya\ used  Geert Hofstede's cultural dimensions~\cite{hoftede2010cultures}, such as individualism index, and showed that users from higher individualism index countries exhibit higher level of concern about their privacy compared to the users from collectivistic countries.
In this study, we focus on understanding how the users' behavior, characterized by broad engagement, accomplishments and deviance metrics, relates to their privacy concerns.
%======================================
\section{Dataset Description}\label{sec:ya-details-datasets}
%======================================

Launched by Yahoo! in 2005, \ya\ is available in 12 languages and has $56$M monthly visitors in U.S. alone\footnote{http://www.listofsearchengines.org/qa-search-engines}.
The functionalities of the \ya\ platform and the dataset used in this analysis are presented next.

%-----------------------------------------------
\subsection{The YA Platform}\label{subsec:platform}
%-----------------------------------------------
%\ainote{things to define/explain:\\
%-- user contributed content: answers, questions, comments;\\
%-- points and the incentive mechanism; user levels and what comes with them;\\
%-- self policing: flags. Who can apply them; what restrictions; how are they validated; user levels vs. validation; editors; consequences\\
%-- social network and update dissemination;\\
%-- 
% many of the users may be bogus or dormant: users who signed up, created a few friends, and disappeared quickly. It may be difficult to predict anything about such users. In order to prevent these bogus
%users from skewing the results of our study, we remove, from our dataset, the users with less than 5 friends across Facebook.
%The size of our compressed dataset is 1, 282, 563. Out of the 679, 351 users who specified their genders, the percentage of males is 52.97%. Table 1 shows the properties of the dataset before and after the elimination of bogus users. In this paper, we do all processing on the reduced data set after elimination of bogus users.}
\ya\ is a CQA platform in which community users ask and answer questions on various topics in  predefined taxonomies, e.g., \emph{Business \& Finance}, \emph{Cooking}, and  \emph{Politics \& Government}.
A question consists of a title and a body (typically, additional details).
Members can find questions by searching or browsing through the hierarchy of categories.
%A question has a title (typically, a short summary of the question), and a body with additional details. 

The goal of asking a question is to find a best answer for the question.
Users can write one answer per question and a question remains open for four days to answer.
The asker can extend the answering duration for an extra four days.
If the asker of the question selects a best answer within this time period, \ya\ archives it as a \emph{reference} question and only comments can be added to a reference question.
The asker can rate a best answer between one to five, which is known as \emph{answer rating}.
However, if the asker doesn't select a best answer, community members get an opportunity to vote for a best answer.
\ya\ deletes all unanswered questions when the answering duration expires.

Users in \ya\ can flag content (questions, answers or comments) that violates the Community Guidelines and Terms of Service using the ``Report Abuse'' functionality.
\emph{YA} requires its users to follow the Community Guidelines that forbid users to post spam, insults, or rants, and the Yahoo Terms of Service~\cite{YahooAnswersCommunity} that limits harm to minors, harassment, privacy invasion, impersonation and misrepresentation, and fraud and phishing. 
%Figure~\ref{fig:abuse_report} shows how reporting is done on a content.
Users click on a flag sign embedded with the content and choose a reason between violation of the community guidelines and violation of the terms of service.
They can select between two reasons: violation of the community guidelines (e.g., chat or rant, adult content, spam, insulting other members, etc.), or violation of the terms of service (e.g., harm to minors, violence or threats, harassment or privacy invasion, impersonation or misrepresentation, fraud or phishing, etc.).
Reported content is then verified by human inspectors before it is deleted from the platform.

There is a point system in \ya\ to encourage and reward participation.
%\emph{YA} has a system of points and levels to encourage and reward participation\footnote{https://answers.yahoo.com/info/scoring\_system}.
In short, a user is given two points for answering a question; ten points for a best answer.
However, the user is penalized five points for asking a question, but if she chooses a best answer for her question, three points are given back.
Users are ranked daily on a \emph{leaderboard} based on their points.
The points are also used to split users into seven levels (e.g., 1-249 points: level 1, 250-999 points: level 2, ..., 25000+ points: level 7).
\ya\ uses the  levels  to limit user actions, such as posting questions, answers, comments, follows, and votes: e.g., first level users can ask $5$ questions and provide $20$ answers in a day.
%A leaderboard, updated daily, ranks users based on the total number of points they collected.
%\ainote{Need to explain the levels here: how many, how are they defined, what are the consequences.}
%\ainote{Any related work that studies points, incentives, etc in YA?}

\ya\ users follow each other and create a  Twitter-like follower-followee relationship.
Users are free to follow anyone.
%Users in \emph{YA} can choose to follow other users, thus creating a follower-followee relationship used for information dissemination. 
The followee's  questions, answers, ratings, votes, best answers and awards are automatically disseminated to the followers' newsfeed. 
%\ainote{is this e.g. or should it be i.e.? that is, if the list within parenthesis is complete, than it should be i.e.}
In addition, users can follow questions, in which case all responses are sent to the followers of that question.
% \ainote{``which is another information shared with that user's followers.'' Should this be maybe:  ``in which case all responses are sent to the followers of that question''}
%A user's followers can also learn whether his answer is selected the best or whether he is following a question.
%\ainote{so, there is a feature for following questions, as well? IK: yes, they can}
Users can control the exposure of their information using two options in the privacy settings:
they can choose to hide their content (questions and answers), and they can also choose to hide their network (followers and followees) from other users.

%follower-followee (FF) directed social graph.
%Users can follow each other using a  ``follow'' functionality from their homepage.
%If a user A is following another user B, then an edge from A to B ($A\rightarrow B$) is generated.
%In this case, A is a follower of B and B is a followee of A.

%-----------------------------------------------
\subsection{Dataset}
%-----------------------------------------------

We studied a sample of 1.5 million users, who were active between 2012 and 2013.
These users are connected via $2.6$ million follower-followee relationships in a social network that has $165,441$ weakly connected components.
The largest weakly connected component has $1.1$M nodes ($74.32$\% of the nodes) and $2.4$M edges ($91.37$\% of the edges).
Our study includes all the users.
%Out of the 1.5 million users, $87.20$\% have kept their QA and network privacy to private (only visible to followers) and $9.74$\% users have kept to public (visible to members or visitors). 
%Among the rest $3.06$\% users, $2.23$\% have kept only the QA privacy to private and $0.81$\% have kept only the network privacy to private.

The available privacy configurations allow 4 user groups:
\begin{enumerate}
\item Public: all information is publicly visible ($87.20$\% of users). This is the default setting. 
\item QA-private: only Q/A information is private ($2.23$\% of users), i.e., their questions and answers are visible only to their followers. 
\item Network-private: only network information is private ($0.81$\% of users), i.e., only their followers see the network. 
\item Private: Q/A and network information is private ($9.74$\% of users), thus only visible to the user's  followers.
\end{enumerate}

In the rest of the study we collectively refer to the QA-private and network-private users as semi-private. 
The default privacy in \ya\ is public.
It might be possible that many of the users in the pubic group are dormant: users who signed up, asked and answered some questions, and disappeared quickly. 
These users might skew the results of our study,
%In order to prevent the dormant users,
thus, we only consider active users, who have asked and answered more than 10 questions. 
The active users are about $68$\% of the population and out of them $84.43$\% are public, $2.50$\% are  QA-private, $0.89$\% are network-private, and $12.16$\% are private.
We note that our observations remain the same even if we consider more active users by filtering-in users who have asked and answered more than 20 questions.

\section{Privacy and User Behavior}\label{sec:privacy}

Our goal is to study the association between privacy concerns and behavior in \ya. % the community question answering platform. 
Previous works~\cite{boyd2010FB,Staddon2012PCT} on Facebook have inferred users' privacy concerns using their self-reported feedback on privacy.
Rather than self-reporting, which is subject to bias~\cite{gosling1998people}, we use modifications on privacy settings as a proxy for privacy concern.
We measure several characteristics of user behavior that are related to CQA such as engagement, retention, accomplishments, abuse reporting, and deviance. 
We ask the following research questions: 
\begin{enumerate}

\item \textbf{Is privacy preference associated with user engagement?} \\
We consider two metrics of user engagement: retention, which measures the average interval time between consecutive user contributions (addressed in Section~\ref{sec:retention}), and social engagement, given by the number of followers and followees (Section~\ref{sec:soc-engagement}). 
This question aims to investigate the pattern identified in survey-based Facebook studies, but using CQA-specific and more nuanced engagement metrics on longitudinal activity traces.

\item \textbf{Do privacy-concerned users contribute differently to the community than public users?}\\
Users contribute by posting questions and answering others' questions. %; and by reporting violations of community rules. 
The quality of user-generated content is measured in the number of best answers and the askers' satisfaction with the answer received.
The overall activity is measured in points. 
We characterize user contributions quantitatively and qualitatively in Section~\ref{sec:accompl}. % and user reporting of abuses in Section~\ref{}. . 

%{\color{blue}{
\item \textbf{Do privacy-concerned users have different perception on answer quality than public users?}
Users can themselves select best answers for their posted questions or they can rely on community voting to mark the best answers.
In Section~\ref{sec:answer-quality}, we look at how the community sees the best answers selected by the users who received them. %answer whether privacy-concerned users are more likely to select best answers by themselves.
Specifically, we compare the quality of best answers selected by privacy-concerned users with those selected by public users in terms of the number of thumbs-up and thumbs-down given by the community.
%}}

\item \textbf{Are privacy-concerned users also more abuse-conscious?}
Intuitively, engagement is also correlated with the desire to keep the community free of unethical users (who, for example, may post spam in violation of the community rules). 
The related analysis is presented in Section~\ref{sec:abuse-reporting}.%Do they report more abuses than others?

%   \item Do privacy-concerned users have higher quantitative and qualitative contributions than public users?
  %   \item Which privacy group users are more socially engaged? 

\item \textbf{Are privacy-concerned users more likely to violate community rules?} 
Intuitively, reduced visibility can give a false sense of confidence that might lead to violations of community rules. 
One study~\cite{blackburn2014cheating} in online gaming social networks shows that  newly found and banned cheaters are more likely to change their profile to a more restrictive privacy settings than non-cheaters.
In \ya, we ask, is  this observed more with privacy-concerned users than public users? 
This question is studied in Section~\ref{sec:deviance}.

%\item \textbf{Can we predict privacy settings?}\\ 
%Given the quantitative behavioral differences we expect to observe from our previous research questions, a natural conclusion would be whether we can use the differences to predict privacy settings of the users in CQA platforms?
%To answer that, we present a classification model in Section~\ref{sec:classification}.

\end{enumerate}

%We split users into four groups (public, QA private, network private, and private) considering the combinations of privacy settings.
%%Intuitively, private users are more concerned about their privacy and they do not want to share their contents and network with all members and visitors of YA.
%%On the other hand, public users are either comfortable sharing information with others, or they are not aware of the privacy settings.

\ignore{
Users whose privacy concern is in between public and private (QA-private and network-private) are expected to be moderately concerned about their privacy.
We refer these users as semi-private later in the analysis and examine whether they show similar characteristics or not.
If there exists any quantitative behavioral difference between private and public users, a reasonable hypothesis would be that the semi-private users' behavior would fall between private and public users' behavior.
}

\ignore{
However, the default privacy in YA is public.
It might be possible that many of the users in the pubic group are dormant: users who signed up, asked and answered some questions, and disappeared quickly. 
These users might skew the results of our study.
In order to prevent these dormant users, we only consider active users, who have asked and answered more than 10 questions. 
Our conclusions remain the same if we consider more active users by filtering users who have asked and answered more than 20 questions. 
}

%---------------------------------
\subsection{Privacy and Retention}\label{sec:retention}
%---------------------------------

We define retention as the inverse of the average time difference between two actions not marked as abusive (i.e., fair).
We consider two types of retention, based on questions and answers. 
For both types, if a user has a high average time difference between two fair actions, her retention is low.

\begin{figure}[ht]%
    \centering
    %\subfloat[]{{\includegraphics[width=3.78cm]{plot/retention/question_retention_ccdf.pdf} }}%
    %\qquad
\includegraphics[width=9cm]{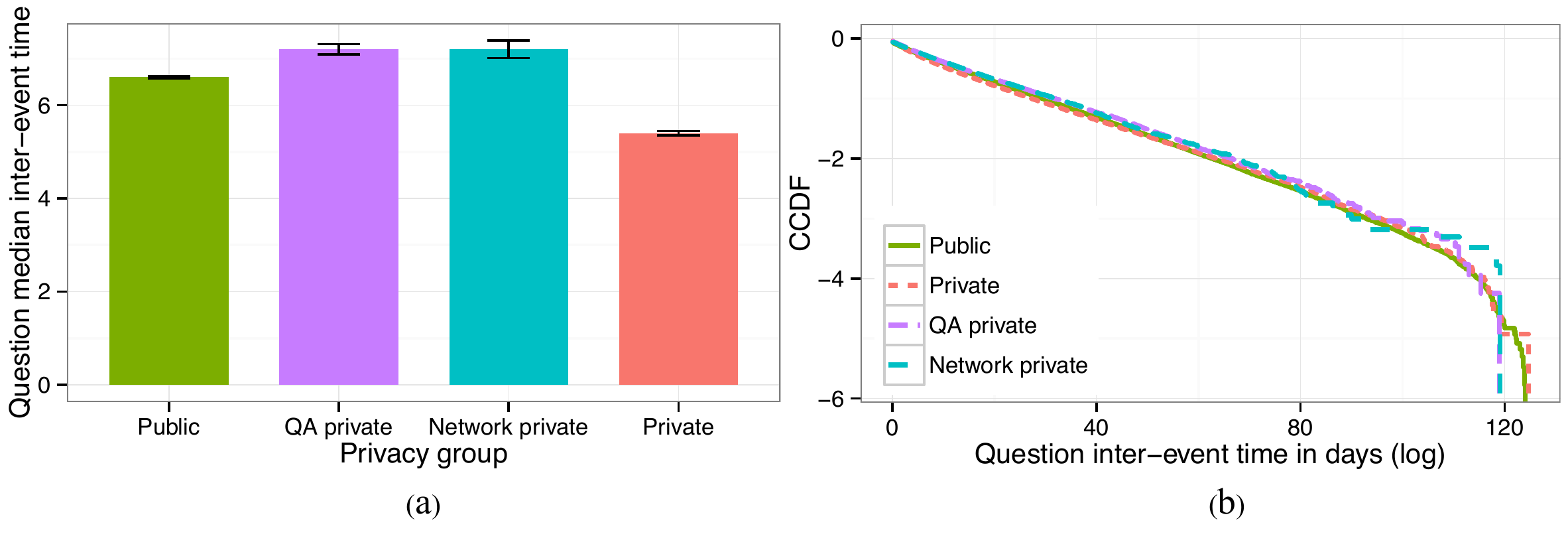}
    \caption{(a) Median of question inter-event time in days with standard error bars; (b) CCDF of question inter-event time in days.}%
    \label{fig:question-retention}%
\end{figure}

\begin{figure}[ht]%
    \centering
    %\subfloat[]{{\includegraphics[width=3.78cm]{plot/retention/question_retention_ccdf.pdf} }}%
    %\qquad
\includegraphics[width=9cm]{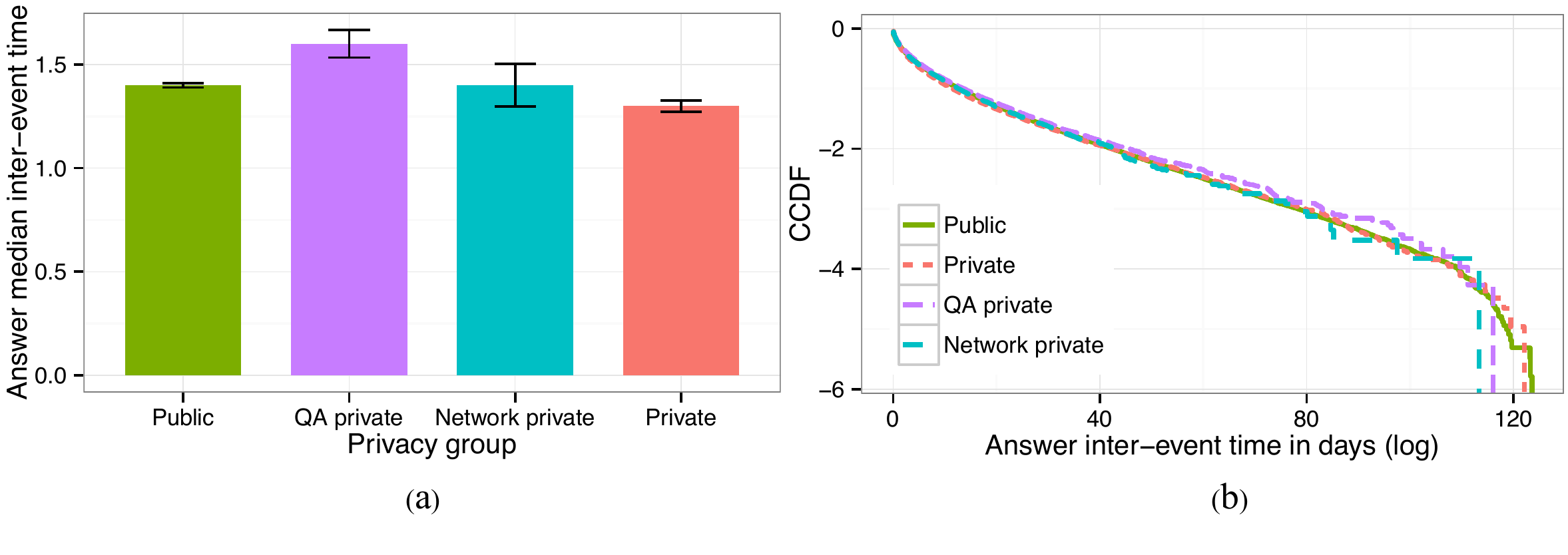}
    \caption{(a) Median of answer inter-event time in days with standard error bars; (b) CCDF of answer inter-event time in days.}%
    \label{fig:answer-retention}%
\end{figure}

Figures~\ref{fig:question-retention}(a)  and (b) show the medians and CCDF of the question inter-event time for the different groups, respectively.
On average, private users have lower question inter-event time (thus higher retention) than public users.
The answer inter-event time in Figures~\ref{fig:answer-retention}(a) and (b) show similar patterns.
It seems semi-private (QA-private and network-private) users have higher average inter-event time, compared to private users, but similar to public users.

We performed a Kruskal-Wallis test to assess the difference  among privacy groups  in terms of retention.
The test shows that at least one of the groups is different from at least one of the others  for question ($\chi ^2 = 458.83, df = 3, p < 2.2e-16$) and answer retention ($\chi ^2 =  119.32, df = 3, p < 2.2e-16$). 
All-pairwise comparison tests after the Kruskal-Wallis test show that besides the QA-private and network-private for question retention and network-private and private for answer retention, all others are different for questions and answers retention $(p $$<$$0.05)$.
These results show that privacy-concerned users are more retained than others.

%\begin{figure}[ht]%
%    \centering
%    %\subfloat[]{{\includegraphics[width=3.78cm]{plot/retention/question_retention_ccdf.pdf} }}%
%    %\qquad
%    \subfloat[]{{\includegraphics[width=6.5cm]{plot/retention/question_retention_boxplot.pdf} }}%
%     \qquad   
%    \subfloat[]{{\includegraphics[width=6.5cm]{plot/retention/answer_retention_boxplot.pdf} }}%
%    \caption{(a) question and (b) answer inter-event time in days with standard error bar.}%
%    \label{fig:question-answer-retention}%
%\end{figure}

%---------------------------------
\subsection{Privacy and Social Circles}\label{sec:soc-engagement}
%---------------------------------

\begin{figure}[ht]%
\centering
\includegraphics[width=9cm]{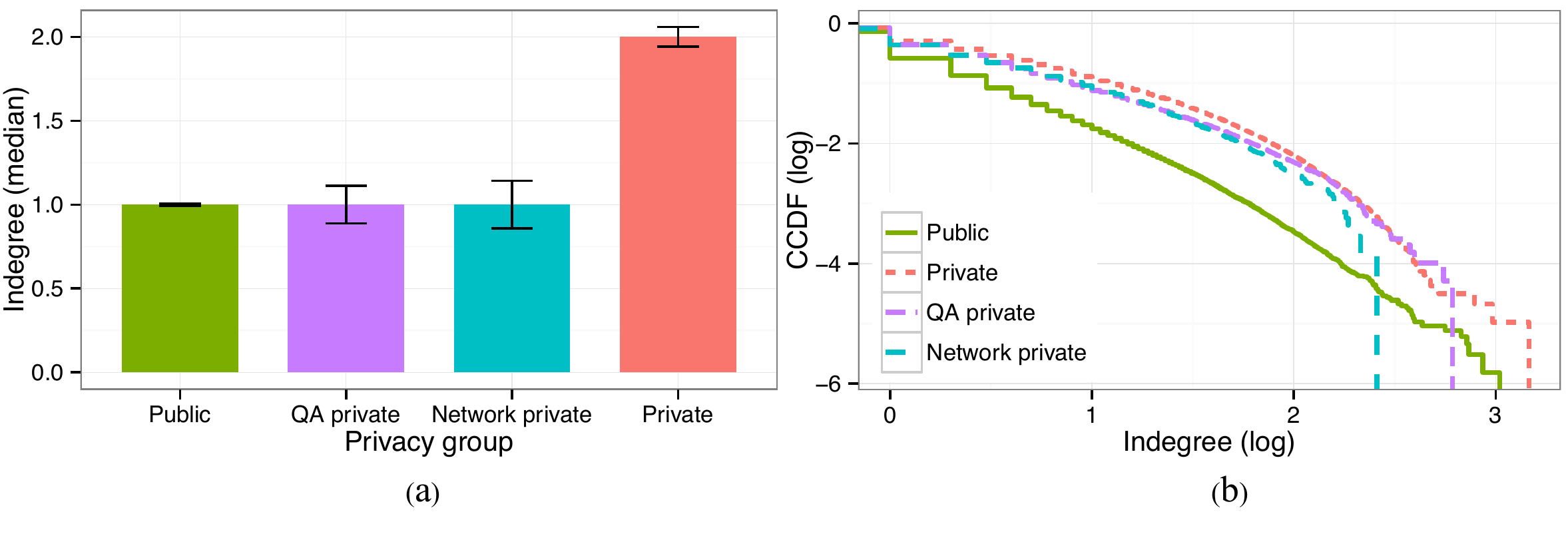}
 \caption{(a) Median of indegree with standard error bars; (b) CCDF of indegree.}%
 \label{fig:indegree}%
\end{figure}

\begin{figure}[ht]%
\centering
\includegraphics[width=9cm]{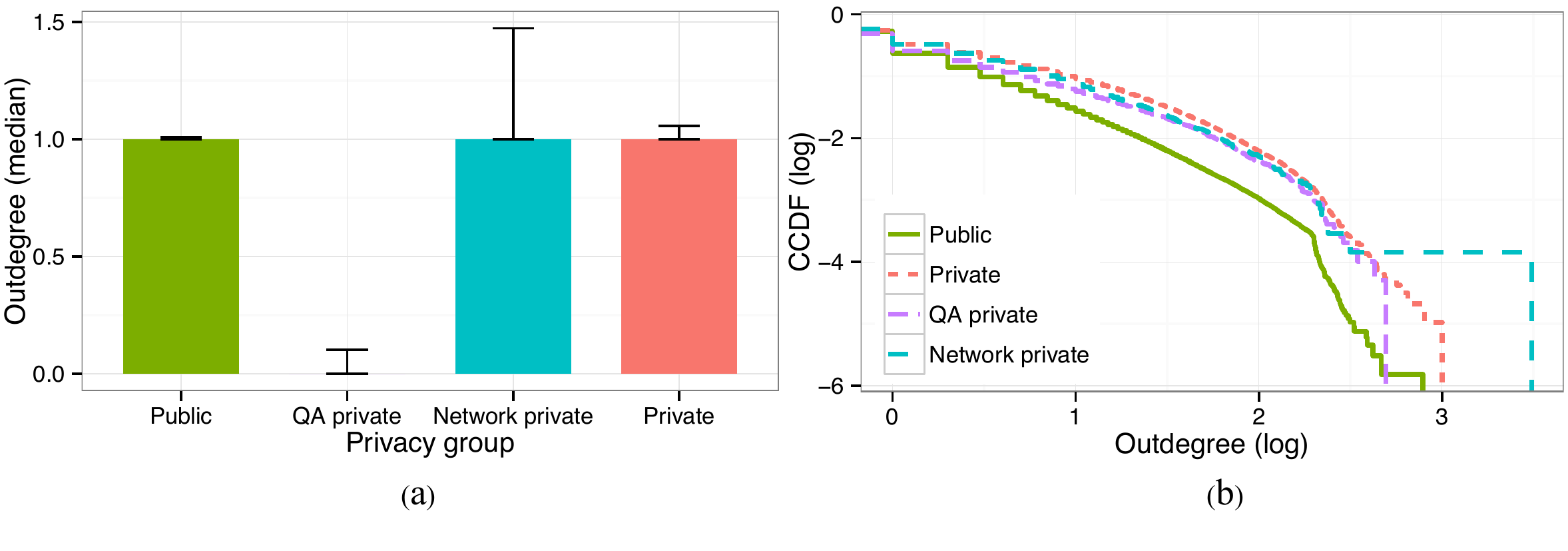}
 \caption{(a) Median of outdegree with standard error bars; (b) CCDF of outdegree.}%
 \label{fig:outdegree}%
\end{figure}

\ya\ users can follow each other, thus, we compute the indegree (total number of followers) and outdegree (total number of followees) of different privacy group users.
Figures~\ref{fig:indegree}(a) and~\ref{fig:outdegree}(a) show the median of indegree and outdegree, respectively, for the four privacy groups.
The CCDF of indegree and outdegree of them are shown in Figures~\ref{fig:indegree}(b) and~\ref{fig:outdegree}(b), respectively.
%Figures~\ref{fig:indegree}(b) and~\ref{fig:outdegree}(b) show CCDF of indegree and outdegree, respectively, for the four privacy groups.
While 20.56\% of private users have more than 5 followers, only 4.42\% of public users do.
However, 15.33\% of network-private and 14.48\% of QA-private users have more than 5 followers.
Alternatively, while 14.79\% of private users follow more than 5 users, only 5.85\% of public users do.
For network-private and QA-private users, these numbers are 12.92\% and 9.79\%, respectively.

The results indicate that more restrictive private settings users have richer social circles.
Indeed, Kruskal-Wallis tests show that at least one of the privacy groups is different from at least one of the other groups, for both the indegree ($\chi ^2 =  29383.67, df = 3, p  < 2.2e-16$) and outdegree ($\chi ^2 =  2913.63, df = 3, p  < 2.2e-16$).
All-pairs comparison tests between the privacy groups show that all pairwise privacy groups are different $(p  < 0.05)$ for indegree, and only network-private and private users are same for outdegree $(p  < 0.05)$.

%\begin{figure}[ht]%
%    \centering
%    \subfloat[]{{\includegraphics[width=6.5cm]{plot/social/indegree_ccdf.pdf} }}%
%    \qquad
%    \subfloat[]{{\includegraphics[width=6.5cm]{plot/social/outdegree_ccdf.pdf} }}%
%   % \qquad   
%    %\subfloat[label 2]{{\includegraphics[width=4cm]{plot/social/indegree_boxplot1.pdf} }}%
%
%    \caption{(a) CCDF of Indegree; (b) CCDF of outdegree.}%
%    \label{fig:indegree}%
%\end{figure}

%---------------------------------
\subsection{Privacy and Accomplishments}\label{sec:accompl}
%---------------------------------

We consider two accomplishments that measure the quantity and quality of user contribution, through the point system described in Section~\ref{subsec:platform}.
Quantity of contribution is measured by the points users earn for their activities.
%Note that YA has a system of points and levels to encourage and reward participation. 
%A user is penalized five points for posting a question, but if she chooses a best answer for her question, three points are given back. A user who posts an answer receives two points; a best answer is worth 10 points.
To measure quality of contribution we use two metrics: Best Answer Percentage (BAP) and Award Rating Percentage (ARP).
BAP is the percentage of a user's answers that are selected as best. %, calculated by the following.
%\begin{equation*}
%\textrm{Best answer percentage}= \frac{\textrm{Number of best answers}}{\textrm{Number of answers}}*100
%\end{equation*}
ARP measures how satisfactory a user's best answers are.
A \ya\ asker can rate a best answer from 1 to 5 to declare how satisfied she is with the answer.
$ARP_{j}$ is the average rating a user $j$ receives for her best answers:
\begin{equation*}
ARP_{j}= \frac{ \sum\limits_{i=1}^{\#\textrm{best answers of j}}\textrm{Award rating for best answer i}} {\textrm{\#Total answers of j * 5 }}*100
\end{equation*}

\begin{figure}[ht]%
\centering
\includegraphics[width=9cm]{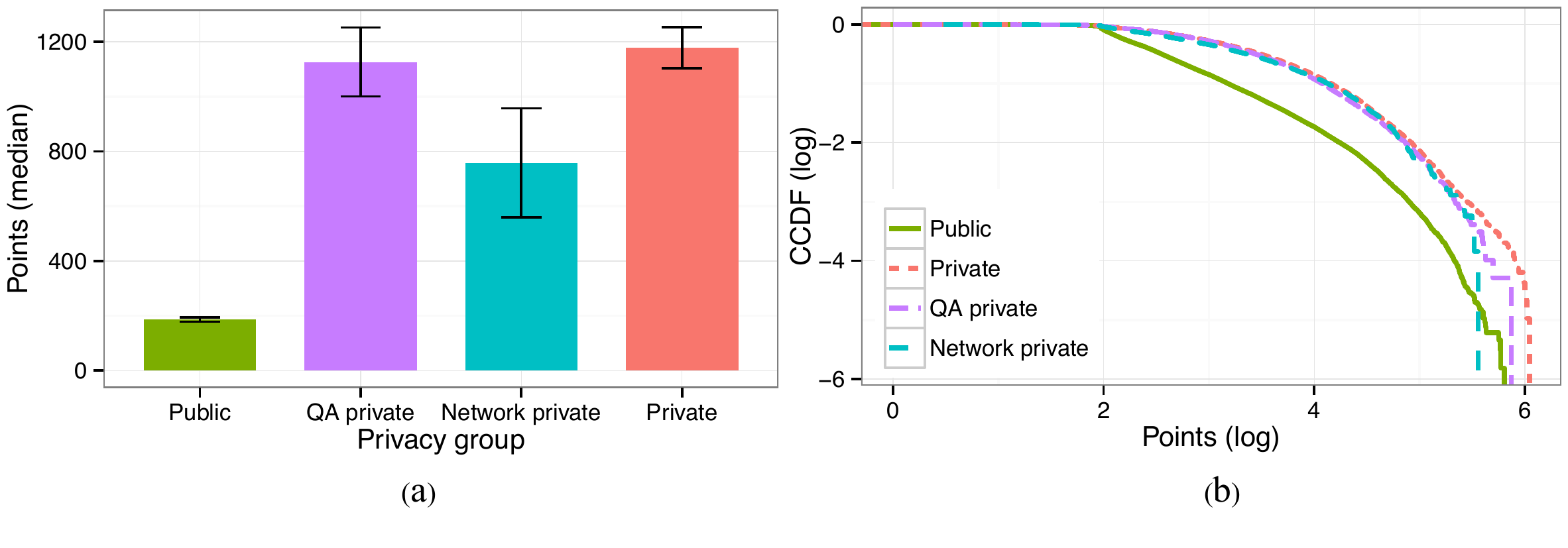}
\caption{(a) Median of points with standard error bars; (b) CCDF of points.}%
\label{fig:points}%
\end{figure}

Figure~\ref{fig:points}(a) shows median points with standard error for different privacy group users.
It appears that median points of private and semi-private users are higher than public users.
In fact, the CCDF of points in~\ref{fig:points}(b) shows that while 53.28\% of private, 52.35\% of QA-private and 45.51\% of network-private users have more than 1000 points, only 14.14\% of public users have more than 1000 points.

A Kruskal-Wallis test shows at least one of the privacy groups is different from at least one of the other groups for  award points ($\chi ^2 =  75884.12, df = 3, p < 2.2e-16$).
Moreover, all-pairs comparison tests between the four privacy groups show that besides private and QA-private, all others are different $(p < 0.05)$.
These results indicate that privacy-concerned users contribute more in YA from a quantitative point of view.

%\begin{figure}[ht]%
%    \centering
%    \subfloat[]{{\includegraphics[width=6.5cm]{plot/accomplishments/points_boxplot.pdf} }}%
%    \qquad
%    \subfloat[]{{\includegraphics[width=6.5cm]{plot/accomplishments/points_ccdf.pdf} }}%
%    % \qquad   
%   % \subfloat[label 2]{{\includegraphics[width=4cm]{plot/accomplishments/points_boxplot1.pdf} }}%
%
%    \caption{(a) Median of points with standard error bar; (b) CCDF of points.}%
%    \label{fig:points}%
%\end{figure}

\begin{figure}[ht]%
\centering
\includegraphics[width=9cm]{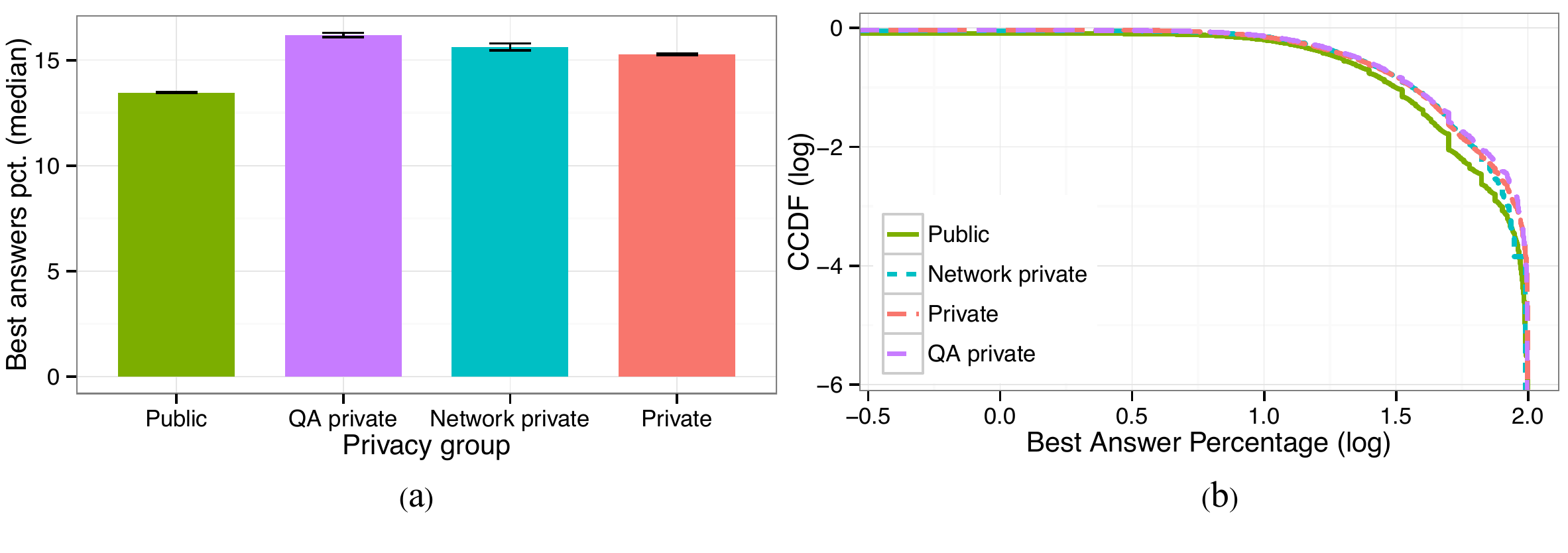}
\caption{(a) Median of best answers percentage with standard error bars; (b) CCDF of best answers percentage.}%
\label{fig:BAP}%
\end{figure}

\begin{figure}[ht]%
\centering
\includegraphics[width=9cm]{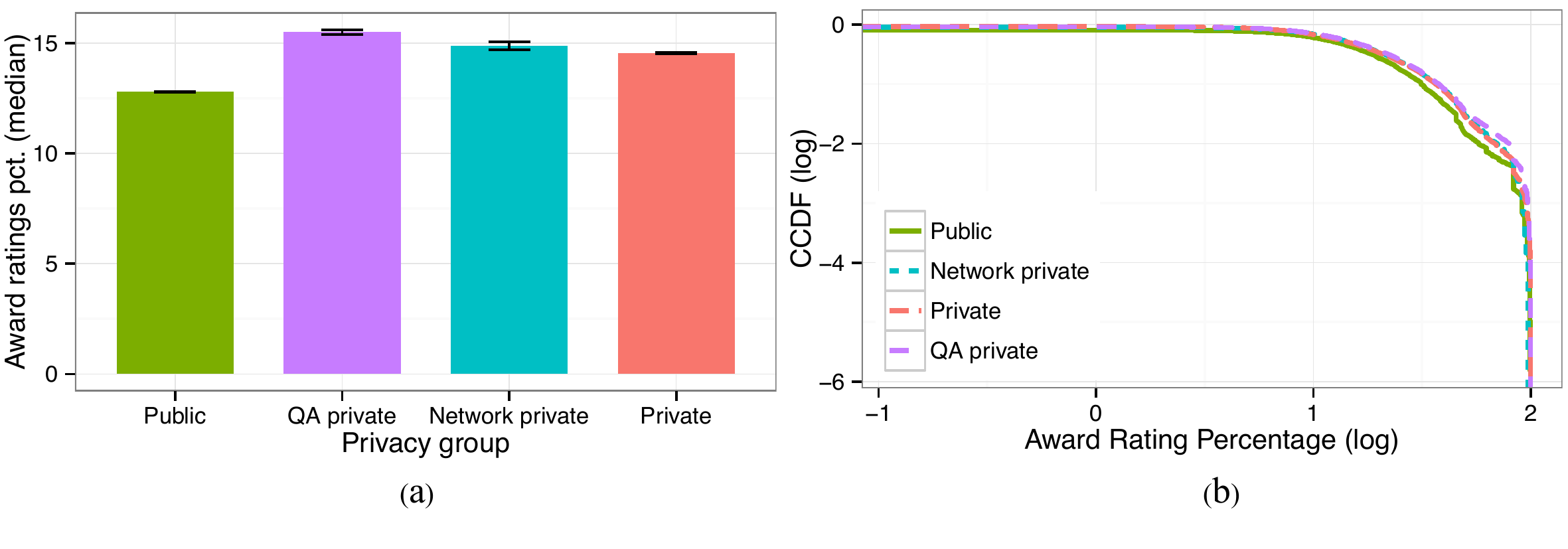}
\caption{(a) Median of award rating percentage with standard error bars; (b) CCDF of award rating percentage.}%
\label{fig:ARP}%
\end{figure}

However, unlike quantitative contributions where public users are far behind the private ones, we found smaller, albeit significant, difference in the qualitative contributions among the four privacy groups.
Figures~\ref{fig:BAP}(a) and~\ref{fig:ARP}(a) show the medians of best answer percentage (BAP) and award rating percentage (ARP) of different privacy group users, respectively.
Although in both cases, private and semi-private group users have higher percentage than public users, the difference is less compared to points (even by a visible inspection on CCDF of BAP (Figure~\ref{fig:BAP}(b)) and ARP (Figure~\ref{fig:ARP}(b)) shows no difference across all privacy groups).
Analyzing the CCDFs we get 27.96\% of public users have best answers percentage more than 20, and 34.91\% of private, 35.42\% of network-private and 37.17\% of QA-private users have best answers percentage more than 20.
On the other hand, 27.10\% of public, 33.56\% of private,  34.03\% of network-private and 36.01\% of QA-private users have award rating percentage more than 20.

For both BAP and ARP, we notice that all privacy groups' numbers (median or CCDF) are close, especially private and network-private.
So, one important question is how different privacy groups are in terms of users' qualitative contribution.
We conducted a Kruskal-Wallis test on both BAP and ARP.
The test results show that at least one of the privacy groups is different from at least one of the other groups for BAP ($\chi ^2 =  5832.93, df = 3, p < 2.2e-16$) and also for ARP ($\chi ^2 =  5604.056, df = 3, p < 2.2e-16$).
Moreover, all-pairs comparison tests between the four privacy groups show that only private and network-private groups are the same $(p < 0.05)$, and all other pairwise privacy groups are different.
Thus, we confirm that privacy-concerned users have higher quantitative and qualitative contributions than others.

%\begin{figure}[ht]%
%    \centering
%%    \qquad    
%    \subfloat[]{{\includegraphics[width=6.5cm]{plot/accomplishments/awardRating_boxplot.pdf} }}%
%     \qquad   
%       \subfloat[]{{\includegraphics[width=6.5cm]{plot/accomplishments/bestAnswer_boxplot.pdf} }}%
%
%   % \subfloat[label 2]{{\includegraphics[width=4cm]{plot/accomplishments/awardRating_boxplot1.pdf} }}%
%
%    \caption{Median of (a) Best answers percentage; (b) Award rating percentage. Standard error bars are also shown.}%
%    \label{fig:BAP}%
%\end{figure}

%\begin{figure}[ht]%
%    \centering
%    \subfloat[]{{\includegraphics[width=6.5cm]{plot/accomplishments/bestAnswer_ccdf.pdf} }}%
%    \qquad
%      \subfloat[]{{\includegraphics[width=6.5cm]{plot/accomplishments/awardRating_ccdf.pdf} }}%
%     %\qquad   
%   % \subfloat[label 2]{{\includegraphics[width=4cm]{plot/accomplishments/bestAnswer_boxplot1.pdf} }}%
%    \caption{CCDF of (a) Best answers percentage; (b) Award rating percentage.}%
%    \label{fig:BAP_CCDF}%
%\end{figure}
%%\vspace{-5mm}

%\begin{figure}[ht]%
%    \centering
%    \subfloat[]{{\includegraphics[width=6.5cm]{plot/accomplishments/bestAnswer_ccdf.pdf} }}%
%    \qquad
%      \subfloat[]{{\includegraphics[width=6.5cm]{plot/accomplishments/awardRating_ccdf.pdf} }}%
%     %\qquad   
%   % \subfloat[label 2]{{\includegraphics[width=4cm]{plot/accomplishments/bestAnswer_boxplot1.pdf} }}%
%    \caption{CCDF of (a) Best answers percentage; (b) Award rating percentage.}%
%    \label{fig:BAP_CCDF}%
%\end{figure}
%%\vspace{-5mm}

%---------------------------------
\subsection{Privacy and Best Answer Quality}\label{sec:answer-quality}
%---------------------------------

In \ya, the best answer of a question is selected either by the asker of the question or by the community.
If an asker does not select the best answer, the community members do that by voting.
We first look at how  different privacy groups are in selecting the best answers by themselves.
We calculate the percentage of the best answers selected out of the total number of questions asked per user. %by the askers with respect to the total number of questions asked per user.

%\begin{figure}[ht]%
%    \centering
%     \includegraphics[width=6cm]{plot/quality/answer_self_selection_ccdf.pdf} 
%    \caption{CCDF of percentage of asker selected best answers.}
%    \label{fig:asker-selected}%
%\end{figure}

%\begin{figure}[ht]%
%    \centering
%    \subfloat[]{{\includegraphics[width=6.5cm]{plot/quality/answer_self_selection_ccdf.pdf} }}%
%    \qquad
%      \subfloat[]{{\includegraphics[width=6.5cm]{plot/quality/answer_self_selection_boxplot.pdf} }}%
%
%     %\qquad   
%   % \subfloat[label 2]{{\includegraphics[width=4cm]{plot/accomplishments/bestAnswer_boxplot1.pdf} }}%
%
%    \caption{(a) CCDF of percentage of asker selected best answers; (b) Median of percentage of asker selected best answers. Standard error bars are also shown.}%
%    \label{fig:asker-selected}%
%\end{figure}

\begin{figure}[ht]%
    \centering
\includegraphics[width=9cm]{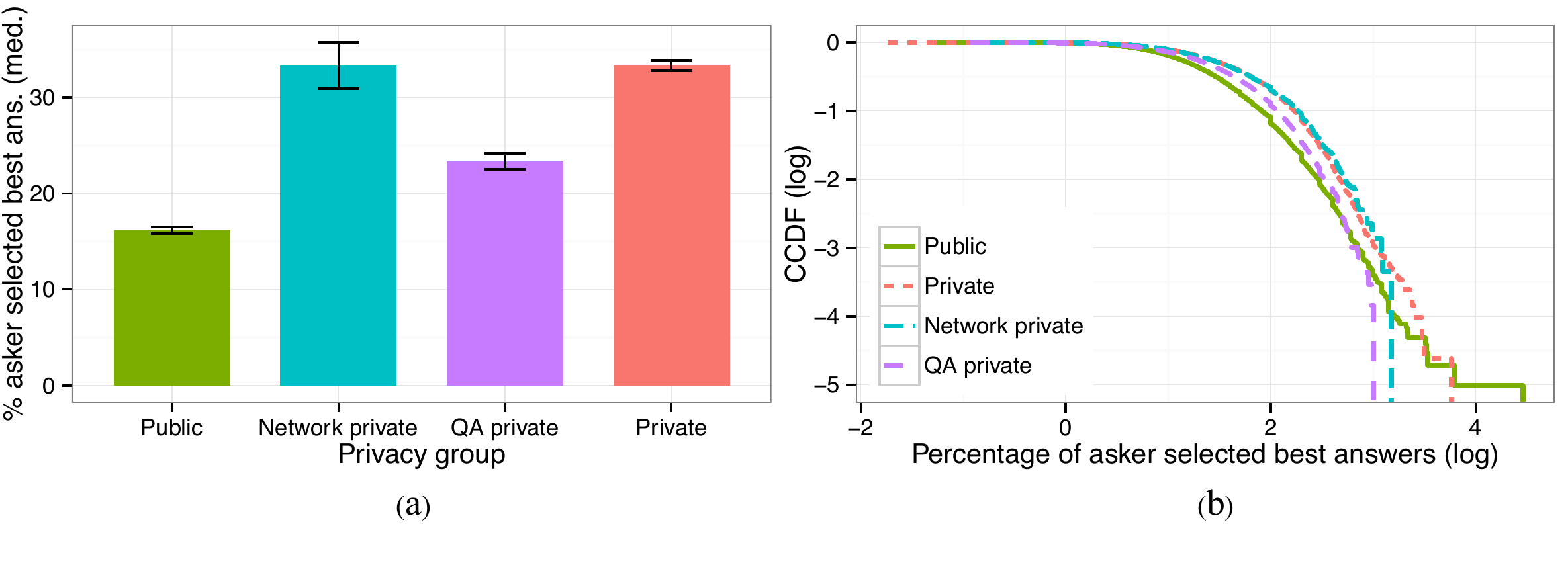} 
    \caption{(a) Median of percentage of asker selected best answers with standard error bars; (b) CCDF of percentage of asker selected best answers.}%
    \label{fig:asker-selected}%
\end{figure}

Figures~\ref{fig:asker-selected}(a)  and  (b) show the median and  CCDF of the percentage of asker-selected best answers  for different privacy group users, respectively. 
Analyzing the distribution, we observe that while 61.35\% of private, 57.85\% of network-private, 51.61\% of QA-private users selected more than 20\% of their best answers by themselves,  only 38.35\% of public users have done the selection by themselves.
A Kruskal-Wallis test shows that at least one of the privacy groups is different from at least one of the other groups in terms of asker-selected best answers ($\chi ^2 =  9522.60, df = 3, p < 2.2e-16$).
All-pairs comparison tests between the four privacy groups show that besides the network-private and private groups, all other pairwise privacy groups are different $(p < 0.05)$.

Next, we focus on the quality of the best answers that users selected by themselves.
We measure this quality based on community members' feedback on those answers.
Community members can provide feedback on answers by giving either a \emph{thumbs up}  or a \emph{thumbs down} (at most one such feedback per answer).
For each user $j$ who selected best answers to his own questions, we calculate the average number of thumbs as the ratio between the positive community feedback and the number of asker-selected best answers.  

\vspace{-3mm}
\begin{equation*}
\textrm{AvgThumbs}_j= \frac{\textrm{\# Thumbs up -- \#Thumbs down}}{\textrm{\# Best answers selected by j}}
\end{equation*}

\begin{figure}[ht]
\centering
\includegraphics[width=9cm]{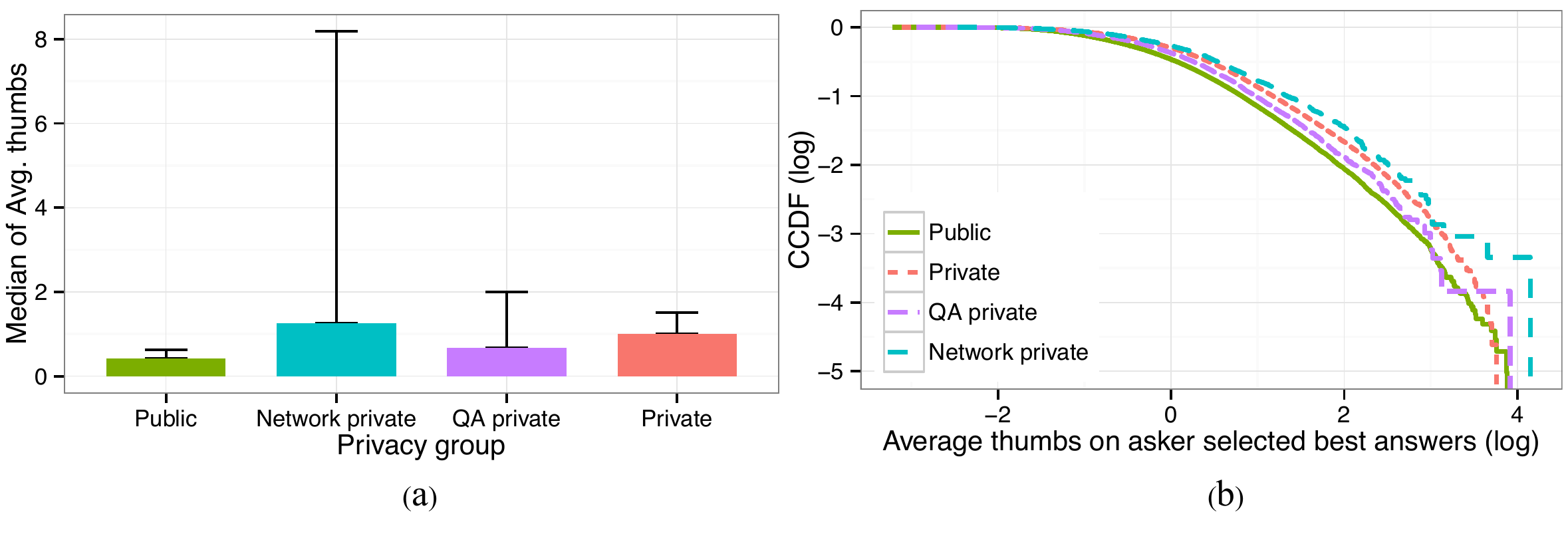} 
\caption{(a) Median of average thumbs on asker selected best answers with standard error bars; (b) CCDF of average thumbs on asker selected best answers.}
\label{fig:average-thumbs}
\end{figure}

Figures~\ref{fig:average-thumbs}(a) and (b) show the median and  CCDF of the average thumbs on best answers selected by the askers, respectively.
The distribution shows that all private group users have more average thumbs on best answers then public users.
We observe that  while 21.40\% of private, 24.62\% of network-private, 16.51\% of QA-private users have got 5 average thumbs on their best answers,  only 11.45\% of public users have got 5 average thumbs on the best answers they selected.
A Kruskal-Wallis test shows that at least one of the privacy groups is different from at least one of the other groups in terms of average thumbs with $\chi ^2 =  5680.47, df = 3, p < 2.2e-16$.
All-pairs comparison tests between the four privacy groups show all  pairwise privacy groups are different $(p < 0.05)$.

%---------------------------------
\subsection{Privacy and Abuse Reporting}\label{sec:abuse-reporting}
%---------------------------------

As a crowd-sourced community, \ya\ relies on its users for self moderation.
%The health of the community depends on community contributions in terms of reporting abuses.
Thus, users not only provide questions and answers, but also report inappropriate content using the abuse report functionality.
If the report is valid, the content is  deleted from the community. 
In this way, users serve as an intermediate layer in the \ya\ moderation process since these abuse reports are verified by human inspectors.
We have already seen that privacy preferences of users have significant association with a number of different dimensions including retention and accomplishments, thus we suspect that privacy is also associated with abuse reporting.

\begin{figure}[ht]%
    \centering
     \includegraphics[width=9cm]{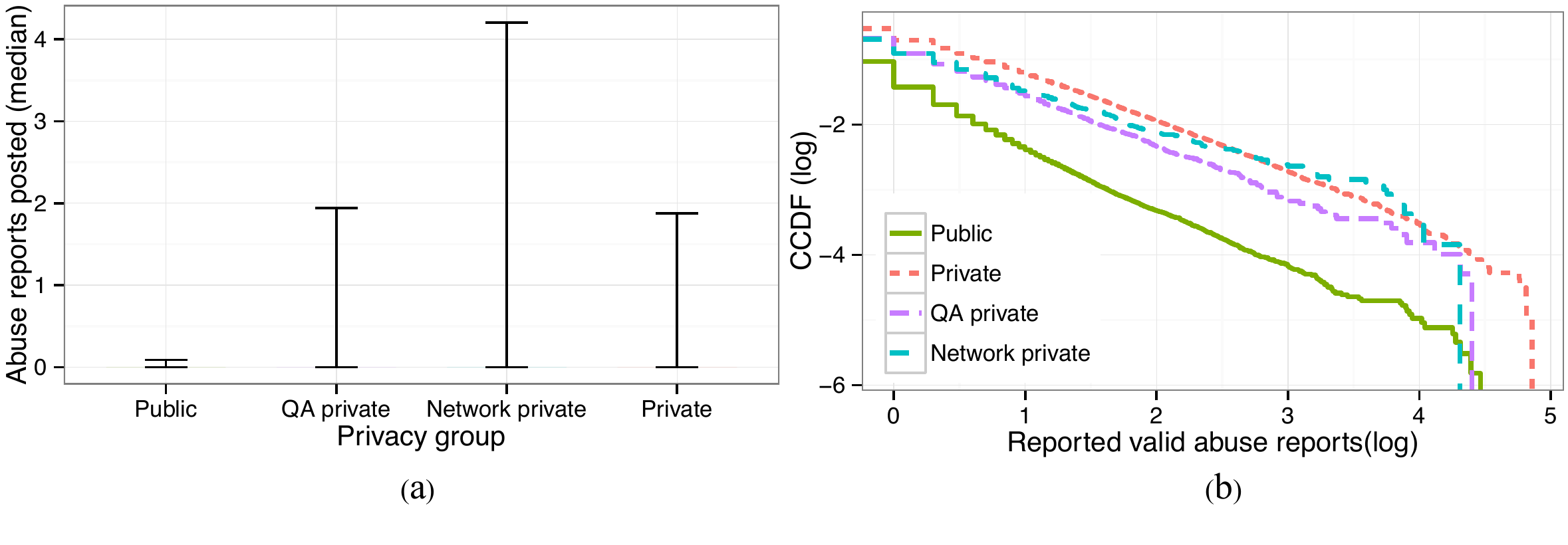} 
    \caption{(a) Median of average valid abuse reports with standard error bars;  (b) CCDF of valid abuse reports.}
    \label{fig:abuseReports}%
\vspace{-5mm}
\end{figure}

The median and CCDF of the valid abuse reports posted by users are shown in Figures~\ref{fig:abuseReports}(a) and (b), respectively.
Although, abuse reports are highly appreciated for maintaining a clean CQA environment, very few people tend to report abuses.
We find that  46\% of the users reported only one abuse and 90\% of abuse reports are contributed by only 7.96\% of users.
So, it's not surprising that all median values are zero in Figures~\ref{fig:abuseReports}(a).
However, the private users have very high variability in abuse reporting compared to the public users. 

The distributions in Figure~\ref{fig:abuseReports}(b) show that, on average, private users have posted more abuse reports than semi-private and public users.
Indeed, all three private groups of users have posted a very large number of valid abuse reports compared to public users.
Analyzing the distribution, we observe that 5.93\% of private, 3.15\% of network-private, 2.73\% of QA-private and only 0.20\% of public users have posted more than 10 valid abuse reports.
A Kruskal-Wallis test shows that at least one of the privacy groups is different from at least one of the other groups in terms of abuse reporting behavior ($\chi ^2 =  37647.77, df = 3, p  < 2.2e-16$).
All-pairs comparison tests between the four privacy groups show that besides the QA-private and network-private groups, all other pairwise privacy groups are different $(p < 0.05)$.

%---------------------------------
\subsection{Privacy and Deviance}\label{sec:deviance}
%---------------------------------

Deviant behavior is defined by actions or behaviors that are contrary to the dominant norms of the society~\cite{douglas1982sociology}.
Although social norms differ from culture to culture, within a context, they remain the same and they are the rules by which the members of the community are conventionally guided. 
\ya\ has established norms as reflected by its community guidelines and terms of service~\cite{YahooAnswersCommunity}.
We define user behaviors as deviant if they depart from these norms.
In our previous work~\cite{Kayes2015WWW} on \ya\ content abusers, we define a \emph{deviance score} metric that indicates how much a user deviates from the norm in terms of received flags considering the amount of the user's activity.
In short, we define the deviance score for a user $u$ as the number of correct abuse reports (flags) she receives over the total content (question/answer) she posted, after eliminating the expected average number of correct abuse reports given the amount of content posted:
\begin{equation}
\centering
\begin{split}
 \textrm{Deviance}_{\textrm{Q/A}}(u)=  Y_{Q/A,u}-\hat{Y}_{Q/A,u} \end{split}
\label{eq:deviance_QA}
\end{equation} 
where $Y_{Q/A,u}$ is the number of correct abuse reports received by $u$ for her questions/answers, and $\hat{Y}_{Q/A,u}$ is the expected number of correct abuse reports to be received by $u$ for those questions/answers.

To capture the expected number of the correct abuse reports a user receives for questions/answers, we considered a number of linear and polynomial regression models between the response variable (number of correct abuse reports) and the predictor variable (number of questions/answers).
Among them, the following linear model was the best in explaining the variability of the response variable.
\begin{equation}
\centering
\begin{split}
 Y= \alpha+ \beta X + \epsilon
 \end{split}
\end{equation} 
where $Y$ is the number of correct abuse reports (flags) received for the content, $X$ is the number of content posts and $\epsilon$ is the error term.
In eq.~(\ref{eq:deviance_QA}), a positive deviance score reflects deviant users, i.e., those whose deviance cannot be only explained by their activity levels.

%\begin{figure}[ht]%
%    \centering
%    \subfloat[]{{\includegraphics[width=6.5cm]{plot/deviance/question_deviance_ccdf.pdf} }}%
%    \qquad
%    \subfloat[]{{\includegraphics[width=6.5cm]{plot/deviance/question_deviance_boxplot.pdf} }}%
%    % \qquad   
%   % \subfloat[label 2]{{\includegraphics[width=4cm]{plot/deviance/question_deviance_boxplot1.pdf} }}%
%    \caption{(a) CCDF of question deviance scores;  (b) Median question deviance scores with standard error bar.}%
%    \label{fig:questionDeviance}%
%\vspace{-5mm}
%\end{figure}

\begin{figure}[ht]%
    \centering
     \includegraphics[width=9cm]{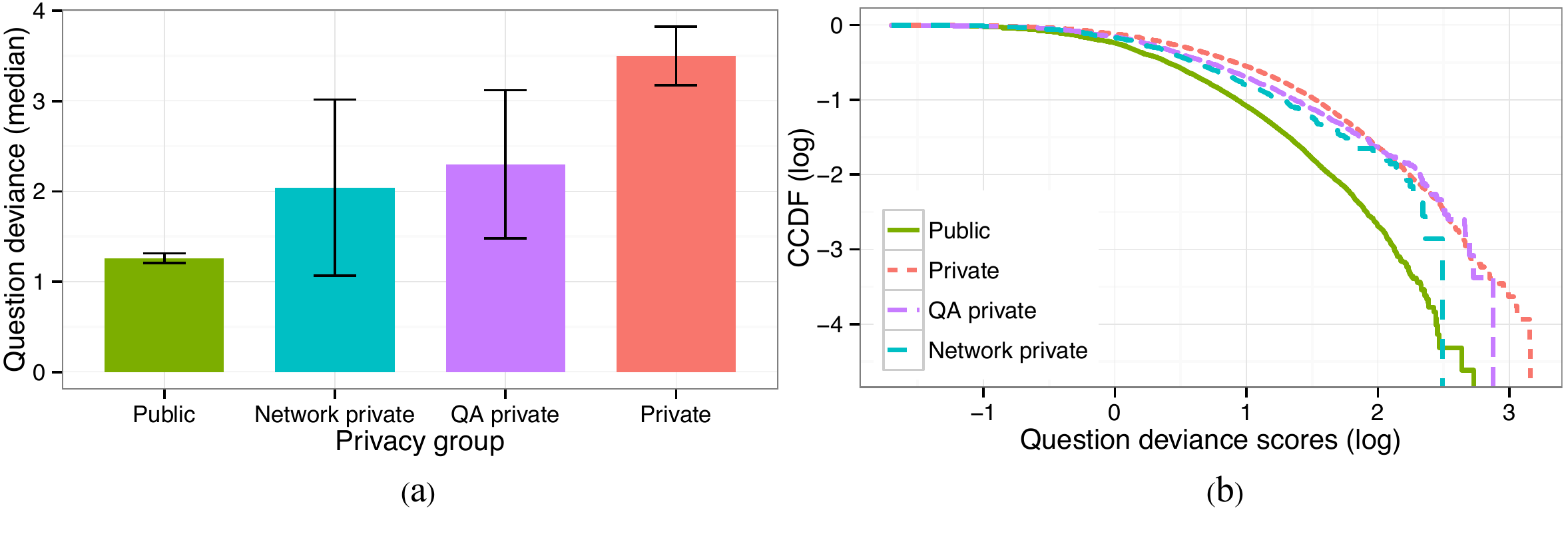} 
    \caption{(a) Median question deviance scores with standard error bars; (b) CCDF of question deviance scores.}%
    \label{fig:questionDeviance}%
\end{figure}

\begin{figure}[ht]%
    \centering
     \includegraphics[width=9cm]{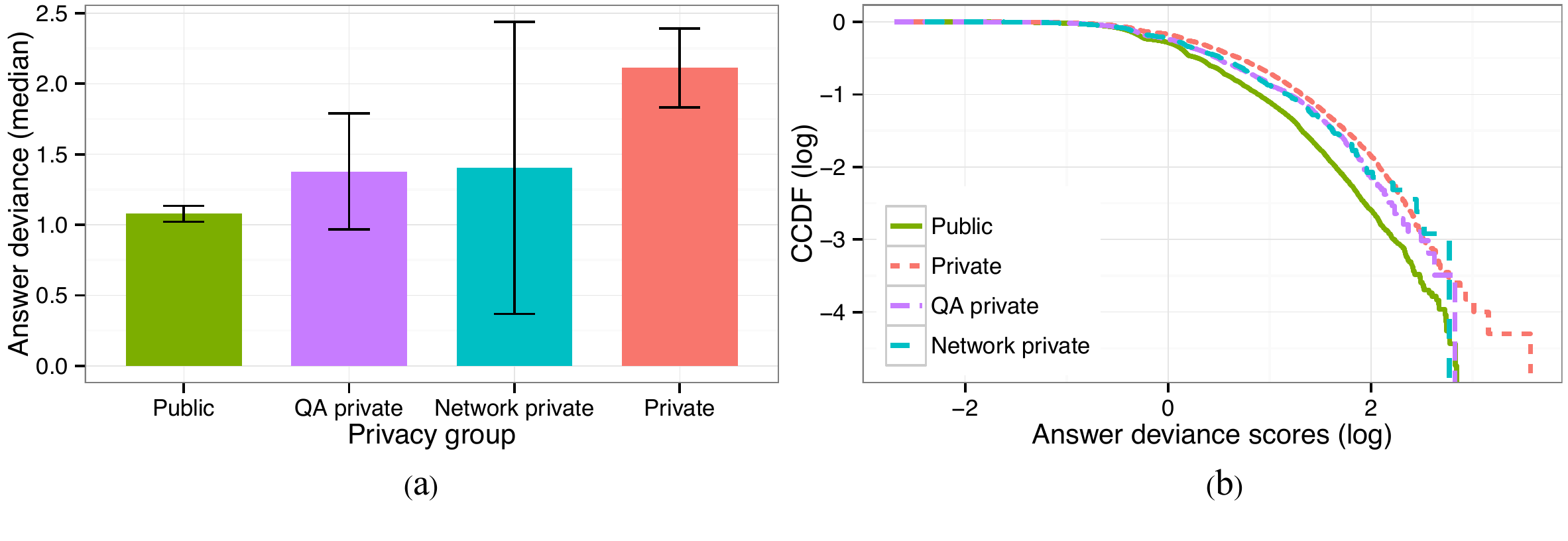} 
    \caption{(a) Median answer deviance scores with standard error bars; (b) CCDF of answer deviance scores.}%
    \label{fig:answerDeviance}%
\end{figure}

%\begin{figure}[ht]
%    \centering
%    \subfloat[]{{\includegraphics[width=6.5cm]{plot/deviance/answer_deviance_ccdf.pdf} }}%
%    \qquad
%    \subfloat[]{{\includegraphics[width=6.5cm]{plot/deviance/answer_deviance_boxplot.pdf} }}%
%     %\qquad   
%    %\subfloat[label 2]{{\includegraphics[width=4cm]{plot/deviance/answer_deviance_boxplot1.pdf} }}%
%    \caption{ (a) CCDF of answer deviance scores;  (b) Median answer deviance scores with standard error.}%
%    \label{fig:answerDeviance}%
%\vspace{-7mm}
%\end{figure}

Figures~\ref{fig:questionDeviance}(a) and (b) show the median and CCDF  of the question deviance scores, respectively.
In both cases, private and semi-private users' question deviance scores are higher than the public users.
Also private users' question deviance scores are higher than semi-private users.
We reach to the same conclusion for the answer deviance scores from the median and CCDF of the answer deviance scores for all users in Figures~\ref{fig:answerDeviance}(a) and (b), respectively.
The Kruskal-Wallis test shows that at least one of the privacy groups is different from at least one of the other groups for question ($\chi ^2 = 4432.72, df = 3, p < 2.2e-16$) and answer deviance scores ($\chi ^2 =  2662.416, df = 3, p < 2.2e-16$).
All-pairs comparison tests between the four privacy groups show that besides the network-private and QA-private groups, all other pairwise privacy groups are different $(p  < 0.05)$.

\section{Summary and Discussions}
\label{sec:discussion}

By performing a large-scale quantitative study, we have shown how users' privacy concerns relate to their behavior in Yahoo Answers, a popular community question answering platform.
We used users' modifications on their privacy settings as a proxy of privacy concerns and grouped users into three main categories: private, semi-private (consisting of two groups, QA-private and network-private), and public.

Our study highlighted a number of results. 
First, we found that $87.20$\% of user accounts on \ya\ are public, the default privacy setting.
This result is similar with Gross and Acquisti's study~\cite{gross2005information} on Facebook, where they found that about 90\% of user profiles maintained the default, public setting.
While expected, this confirmation warns again about the importance of correct default settings (e.g., privacy as contextual Integrity~\cite{kayes13aegis,imrul13out_of_the_wild}) in online applications. 

Second, we discovered that users with enabled privacy settings are more engaged with the community: they have higher retention, more social network contacts, they are better citizens in terms of reporting abuses, overall they contribute more and better content, and have higher perception on answer quality.
This is in line with Staddon et al.'s study~\cite{Staddon2012PCT} on Facebook, who found that users reporting more control and comprehension over privacy are more engaged with the platform.
Therefore, this result is important for two reasons: it applies to a type of online community not previously studied, and it is based on user logs instead of user surveys, prone to self-reporting bias.

Third, we found that, on average, privacy-concerned users show more behavioral deviation in asking and answering questions than users with public accounts.
At a first look, this result seems counterintuitive, given that privacy-concerned users keep the environment clean by reporting more abuses.
However, this result is consistent with our previous study~\cite{Kayes2015WWW}, which finds that deviance in CQA platforms is not necessarily bad.
Deviant users in \ya\ are found to promote user engagement by attracting more users to answer more of their questions.

In addition to characterizing the association between privacy concerns and user behavior, %and the validation of previous, survey-based studies in other online platforms, 
our results may lead to improvements in CQA platforms operation.
Whether an expression of privacy awareness or Internet savviness, users who modify their default privacy settings can be expected to be better citizens. 
If they change their account settings early on in their interaction with the platform, they send a clear signal to platform operators of likely commitment.

CQA platforms could benefit by targeting these users in a number of ways.
For example, the indication of changing privacy settings can be used in \textit{question recommendation}, where questions are routed to the most appropriate users who are more likely to answer.
To find such answerers, typical factors considered are followers, interests, question category, diversity and freshness; privacy settings can also serve as a complementary factor.
Also, some of these users could be assigned community moderating duties to monitor community health, as our results show that they report more abuses.
However, users who do not change their privacy settings are found to be less engaged.
For these users, CQA platforms could provide extra incentives for participation and increased retention.

%who then can route questions accordingly, provide extra incentives for participation, monitor for potential information overflow, etc. 

Our work also shows the importance of user-friendly and more practical design of privacy controls, as we find that increased engagement is associated with the use of privacy controls.
For example, the lack of appropriate visual feedback has been identified as one of the reasons of the under-utilization of privacy settings~\cite{Strater2008StrategisandStruggles}.
A better interface for setting privacy controls in the CQA platforms can impact users' understanding of privacy settings and thus their success in exercising privacy controls.

We acknowledge that our study is observational, hence we can only associate privacy concerns with user behavior. 
In the absence of controlled experimental ground truth data, we cannot draw causal conclusions regarding whether users'  privacy concerns lead to different behavioral pattens in contribution.
Understanding what makes users who change their default privacy settings on a CQA platform to also be more engaged in that community is among our future research objectives. 
The behavioral differences we have found in this paper could be used to create ``privacy recommendation'' in CQA sites, similar to the work of Li et. al~\cite{Li2011Privacy}.
Our future work includes using machine learning techniques on selected behavioral attributes from this study to predict and recommend privacy on Yahoo Answers.

%Moreover, they report more violations of community, contributing thus to maintaining a healthier online community.
%This can be a direct expression of their better engagement in the community: given their larger social circles and their more frequent participation (as shown by higher retention rates), these users get more updates and are thus exposed to more information than the public users, who tend do be more passive. 

%The behavioral differences we have found in this paper could be used to create ``privacy recommendation'' in CQA sites, similar to the work of Li et. al~\cite{Li2011Privacy}.
%Our future work includes using machine learning techniques on selected behavioral attributes from this study to predict and recommend privacy on YA.

%\input{sections/classification}
%\section*{Acknowledgment}
%The work was funded by the National Science Foundation under the  grant CNS 0952420, and by the Yahoo's Faculty Research and Engagement Program.

\bibliographystyle{IEEEtranS.bst}
\bibliography{Bibtex}  % 

\end{document}